\documentclass[11pt,twoside]{article}
\usepackage{asp2010}

\resetcounters

\begin{document}

\title{Variable X-ray absorption in the mini-broad absorption line quasar PG~1126-041}
\author{Margherita Giustini,$^{1,2}$ Massimo Cappi,$^1$ George Chartas,$^3$ Mauro Dadina,$^1$ Mike Eracleous,$^{4,5}$ Gabriele Ponti,$^6$ Daniel Proga,$^7$ Francesco Tombesi,$^{8,9}$ Cristian Vignali,$^2$ and Giorgio G.C. Palumbo$^2$
\affil{$^1$INAF-Istituto di Astrofisica Spaziale e Fisica cosmica di Bologna, via Gobetti 101, I-40129, Bologna, Italy}
\affil{$^2$Dipartimento di Astronomia, Universit\`a degli Studi di Bologna, via Ranzani 1, I-40127, Bologna, Italy}
\affil{$^3$Department of Physics and Astronomy, College of Charleston, Charleston, SC 29424, USA}
\affil{$^4$Department of Astronomy and Astrophysics, the Pennsylvania State University, 525 Davey Lab, University Park, PA 16802, USA}
\affil{$^5$Center for Gravitational Wave Physics, the Pennsylvania State University, University Park, PA 16802, USA}
\affil{$^6$School of Physics and Astronomy, University of Southampton, Highfield, Southampton, SO17 1BJ, UK}
\affil{$^7$Department of Physics and Astronomy, University of Nevada Las Vegas, 4505 Maryland Pkwy Las Vegas, NV 891541-4002, USA}
\affil{$^8$X-ray Astrophysics Laboratory, NASA/Goddard Space Flight Center, Greenbelt, MD 20771, USA }
\affil{$^9$Department of Astronomy and CRESST, University of Maryland, College Park, MD 20742, USA}
}

\begin{abstract} We present the results of a multi-epoch observational campaign
on the mini-broad absorption line quasar (mini-BAL QSO) PG 1126-041 performed
with \textit{\textit{XMM-Newton}} from 2004 to 2009. Time-resolved X-ray spectroscopy and
simultaneous UV and X-ray photometry were performed on the most complete set of
observations and on the deepest X-ray exposure of a mini-BAL QSO to date.
Complex X-ray spectral variability, found on time scales of both months and
hours, is best reproduced by means of variable and massive ionized absorbers along
the line of sight. In the highest signal-to-noise
observation, highly-ionized X-ray absorbing material outflowing much faster
($\upsilon_{out}\sim 16500$ km s$^{-1}$) than the UV absorbing one
($\upsilon_{out}\sim 5000$ km s$^{-1}$) is detected. This highly-ionized
absorber is found to be variable on very short time scales of a few hours.
\end{abstract}

\section{Introduction}

Accretion disk winds are among the most promising physical mechanisms able to
link the small and the large-scale phenomena in active galactic nuclei (AGN), to
shed light on the physics of accretion/ejection around supermassive black holes
(SMBHs), and to help understanding the impact of the AGN phase on the host galaxy
evolution \citep[e.g.][]{1998A&A...331L...1S, 2004ApJ...616..688P, 2005Natur.433..604D,
2009MNRAS.398...53B}.

Such winds are currently directly observed, as blueshifted and broadened
absorption lines in the UV and X-ray spectra of a substantial fraction of AGN.
In the UV band we observe broad absorption lines (BALs, FWHM $> 2000$ km s$^{-1}$),
mini-broad absorption lines (mini-BALs, 500 km s$^{-1}$ $<$ FWHM $<$ 2000 km s$^{-1}$), and
narrow absorption lines (NALs, FWHM $< 500$ km s$^{-1}$) in about 30\% of AGN, and
their intrinsic fraction is estimated to be as high as 50\% \citep{2008MNRAS.386.1426K,
2008ApJ...672..102G, 2011MNRAS.410..860A}. These absorbers can be
outflowing with speeds as high as $\sim 0.2c$. In the X-ray band, we observe
lower-velocity (100-1000 km s$^{-1}$) ionized gas (the so-called ``warm absorber'') in 50\%
of AGN \citep{2005A&A...432...15P, 2007MNRAS.379.1359M}. X-ray high-velocity
absorbers outflowing with speeds up to $\sim 0.3c$ (called ultra-fast outflows, UFOs) are
observed in about 30-40\% of local AGN \citep{2010A&A...521A..57T}. X-ray BALs,
outflowing with speeds up to $0.7c$, have been also observed in a handful of AGN 
\citep{2002ApJ...579..169C,2009ApJ...706..644C,2011xru..conf..240L}. These observational results indicate that
outflowing matter is a key ingredient of the inner regions of
AGN, and that understanding the launching and structure of these winds is
fundamental for constructing complete models of the central engine.

Here we present the main results of a multi-epoch \textit{XMM-Newton}
observational campaign on PG 1126-041, a low-redshift ($z=0.06$) AGN that shows
UV mini-BALs blueshifted by 2000-5000 km s$^{-1}$ in the C~IV, N~V, and Si~IV
species \citep{1999MNRAS.307..821W}. An early \textit{ROSAT} observation suggested the
presence of ionized absorption as well as possible short-term variability
\citep{1999MNRAS.307..821W,2000A&A...354..411K}. The new \textit{XMM-Newton}
campaign provided the largest dataset (a total of four pointings: one archival
in 2004, two in 2008, one in 2009) and the deepest X-ray exposure (133 ks in the
2009 observation) ever on a mini-BAL QSO. Our analysis confirmed the presence of
massive ionized absorbers along the line of sight, and revealed complex spectral
variability on time scales of both months and hours. We focus here on the
variability of the X-ray absorbers; details on the X-ray analysis can be found in
\citet{2011A&A...536A..49G}.

\section{Complex X-ray spectral variability}
\begin{figure}[h!]
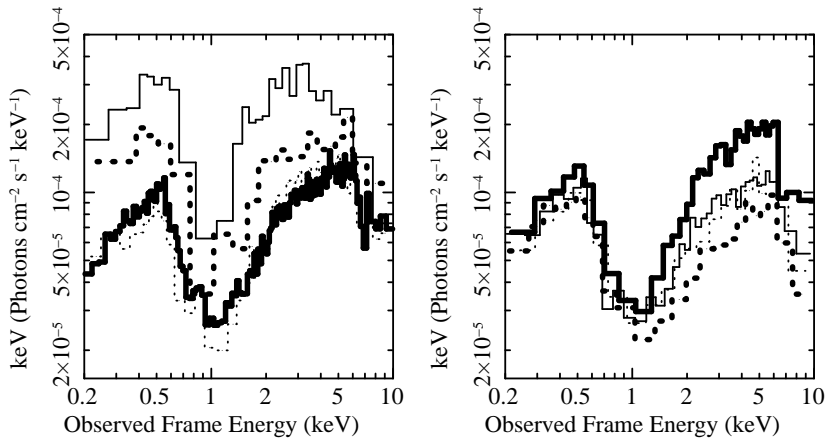
\begin{center}
 \includegraphics[angle=-90,width=5.5cm]{giustini_fig1.ps} 
 \includegraphics[angle=-90,width=5.5cm]{giustini_fig2.ps}\\
 \caption{\label{FIG1}  Left panel: unfolded $F_E$ spectra of PG 1126-041, resolved on time scales of months;
 Dec. 2004 (thin dotted line), June 2008 (thick dotted line),
 Dec. 2008 (thin solid line), and June 2009 (thick solid line) spectrum. Right panel: spectra 
 resolved on time scales of hours;  four contiguous 
 spectral slices of the 2009 long observation are plotted: first (21 ks, thin dotted line), second (30 ks, thick dotted line),
 third (24 ks, thin solid line), and fourth (17 ks, thick solid line) time interval. 
 } 
\end{center}\end{figure}
The mini-BAL QSO PG 1126-041 shows complex and variable X-ray spectral
properties. The average 0.2-10 keV spectra as seen by the EPIC-pn instrument
aboard \textit{XMM-Newton} during the four pointed observations 
are shown in the left panel of Fig.~\ref{FIG1}: most of the spectral
variability on time scales of months occurs at $E<6$ keV. The long June 2009
observation (92 ks of contiguous net exposure time) has been split in four
consecutive time intervals of 21, 30, 24, and 17 ks, where flux variations were
evident. Spectra extracted from each time slice are plotted in the right panel of
Fig.~\ref{FIG1}: strong spectral variability on time scales of hours
affects only the $E>1.5$ keV part of the spectrum. It is evident that two
distinct components are affecting the appearance and the variations of the X-ray
spectrum of  PG 1126-041.

In all epochs, the spectral shape is peculiar and obviously
deviates from a simple power-law. A broad absorption feature is evident
in all the four observations at $E\sim 0.6-1.5$ keV, i.e. the energy range where
X-ray warm absorbers mostly affect the spectral shape of AGN. Indeed a
high-column density, moderately-ionized absorber is detected in every
observation of PG 1126-041. The absorber column density ranges from a
minimum of $N_{m.i.}=3.2^{+0.7}_{-0.4}\times 10^{22}$ cm$^{-2}$ (Dec. 2008
observation), to a maximum of $N_{m.i.}=1.5\pm{0.2}\times 10^{23}$ cm$^{-2}$ 
(June 2009 observation; errors at 3$\sigma$ confidence level). Given the low spectral resolution of the EPIC-pn
instrument, the gas ionization state is poorly constrained: $\log\xi_{m.i.} =
1.55\pm 0.15$ erg cm s$^{-1}$ including the systematic uncertainties of the
model parameters, therefore its possible variations on time scales of months can not
be tracked. As a consequence of the X-ray absorption variability, the observed
optical-to-X-ray spectral index  is
highly variable, going from $\alpha_{\rm{ox}}=-1.7$ in the Dec. 2008
observation, to $\alpha_{\rm{ox}}=-2$ in the June 2009 observation.

\begin{figure}\begin{center}
 \includegraphics[width=11.4cm]{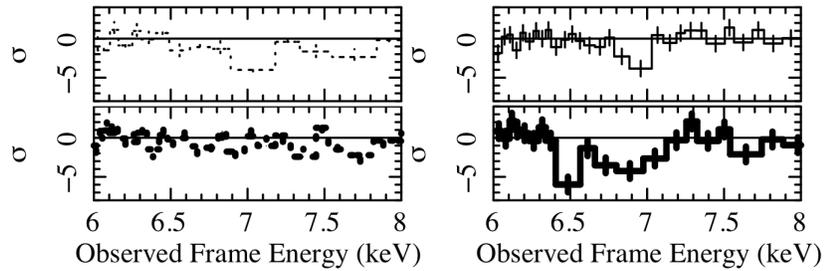}\caption{\label{FIG2} The 6-8 keV EPIC-pn spectral residuals in unit of $\sigma$
 of the four consecutive time slices of the long 2009 observation: first (thin dotted line), second (thick dotted line),
third (thin solid line), and fourth (thick solid line) time interval.} 
\end{center}\end{figure}

A highly-ionized absorber is clearly detected in the iron K band of PG 1126-041
during the long June 2009 observation. In particular, two deep absorption
features at rest frame energy $E\sim 7$ and $\sim 7.4$ keV are identified with
Fe XXV He$\alpha$ and Fe XXVI Ly$\alpha$ transitions, blueshifted by $0.055c$
\citep[see Fig.~4 of][]{2011A&A...536A..49G}. The best-fit parameters for the highly-ionized
outflowing absorber detected in the average 2009 spectra are $N_{h.i.}\sim
7.5\times 10^{23}$ cm$^{-2}$, $\log\xi_{h.i.}\sim 3.4$ erg cm s$^{-1}$, and
$\upsilon_{out}\sim 16500$ km s$^{-1}$. Most interestingly, the highly-ionized
absorber is found to be variable on very short (a few ks) time scales. Fig.~\ref{FIG2} shows the spectral residuals
in the iron K band referred to the four time slices of the 2009 spectrum to a
model where the highly-ionized absorber was removed. Absorption in the iron K
band is significant during the first $\sim 20$ ks of the observation, disappears
during the second time interval, reappears during the third, and then developes
a complex and deep shape during the last $\sim 20$ ks of the exposure, together
with a strong continuum flare. 

The main driver of the PG 1126-041 spectral variability on month
time scales is the variable column density of the moderately-ionized absorber
($\Delta N_{m.i.}/\Delta t \sim 10^{22}$ cm$^{-2}/$month). The intensity of the
intrinsic power-law emission is also found to contribute to the
observed spectral variability, while the photon index $\Gamma\sim 2$ is found to
stay constant within errors between epochs. As for the spectral variability on
time scales of hours, the constant flux at $E\lesssim 1.5$ keV observed during
the long June 2009 observation rules out variability of the moderately-ionized
absorber. On the other hand, the highly-ionized outflowing absorber is
found to be significantly varying over time scales of hours; however, it was not possible to assess
whether it varied in ionization state, column density, or blueshift. The short time scale variability suggests that
the absorber is very compact and close to the X-ray source, and that we are
observing rapid mass ejections from the inner regions of
the accretion disk. In any case, the observed short-term X-ray spectral variability of PG
1126-041 is dominated by variations of the intensity of the intrinsic continuum.


\section{Conclusions}

The \textit{XMM-Newton} observational campaign on the mini-BAL QSO PG 1126-041
provided new results about the link of the accretion and ejection processes in
AGN. Time-resolved X-ray spectroscopy performed on long and short time scales
revealed the presence of massive and variable ionized absorbers along the line
of sight, as well as a varying intensity of the intrinsic X-ray continuum. In
particular, the long 2009 observation provided some of the best evidences of the
transient character of high-velocity, highly ionized X-ray absorbing outflows
over very short time scales of only a few hours.

The appearance of the X-ray spectrum of the mini-BAL QSO PG
1126-041 is dominated by variable ionized absorption. The variability of the
moderately-ionized absorber is qualitatively consistent with the scenario
depicted by \citet{2010ApJ...719.1890M}, where a variable X-ray absorbing screen is
invoked to explain the observed variability over (rest frame) week/month time
scales of the UV mini-BALs in the QSO HS 1603+3820. Our observational
findings are also consistent with UV line-driven accretion disk winds
theoretical scenarios, where a thick, variable screen of X-ray absorbing gas (an
overionized ``failed wind'') close to the central SMBH shields the more distant
portion of the flow from the strong ionizing continuum, and allows for its
effective acceleration \citep[e.g.][]{2004ApJ...616..688P}. Indeed, the observed X-ray
spectra of PG 1126-041 are quite similar to those predicted
for highly-accreting AGN by the hydrodynamical simulations of \citet{2004ApJ...616..688P},
as developed by \citet{2010MNRAS.408.1396S}.

The strong variability of the
highly-ionized outflowing absorber suggests that the assumption of
constant velocity, usually used to estimate the mass outflow rate associated to the
highly-ionized part of AGN winds, is likely to be incorrect: the dynamics of the inner accretion/ejection
flow around SMBHs needs to be further investigated and, to this end,
time-resolved X-ray spectroscopy is the most powerful tool that we have.
Many exciting questions are still open: what is driving the variability of the
ionized absorbers over different time scales? Is there any relation with the
X-ray continuum flux level? What are the duty cycles of the ionized absorbers?
Most importantly, how all this reflects on the mass ouflow rate and on the energy budget associated to the
wind? These questions may be tackled by means of long and deep X-ray
monitoring campaigns of bright BAL and mini-BAL QSOs. 

\acknowledgements MG, MC, MD, and CV acknowledge financial support from the
ASI/INAF contract I/009/10/0. GC and MG acknowledge support
provided by NASA grant NNXlOAEllG. DP and MG acknowledge support provided by the
Chandra award TM0-11010X issued by the CXC, which
is operated by the SAO for and on behalf of
NASA under contract NAS 8-39073. ME acknowledges support from the NSF
under grant AST-0807993. GP acknowledges support via an EU
Marie Curie Intra-European Fellowship under contract no.
FP7-PEOPLE-2009-IEF-254279. 
\bibliographystyle{asp2010}
\bibliography{mg}

\begin{thebibliography}{}
\expandafter\ifx\csname natexlab\endcsname\relax\def\natexlab#1{#1}\fi
\expandafter\ifx\csname url\endcsname\relax
  \def\url#1{\texttt{#1}}\fi
\expandafter\ifx\csname urlprefix\endcsname\relax\def\urlprefix{URL }\fi
\providecommand{\eprint}[2][]{\url{#2}}

\bibitem[{{Allen} et~al.(2011){Allen}, {Hewett}, {Maddox}, {Richards}, \&
  {Belokurov}}]{2011MNRAS.410..860A}
{Allen}, J.~T., {Hewett}, P.~C., {Maddox}, N., {Richards}, G.~T., \&
  {Belokurov}, V. 2011, \mnras, 410, 860. \eprint{1007.3991}

\bibitem[{{Booth} \& {Schaye}(2009)}]{2009MNRAS.398...53B}
{Booth}, C.~M., \& {Schaye}, J. 2009, \mnras, 398, 53. \eprint{0904.2572}

\bibitem[{{Chartas} et~al.(2002){Chartas}, {Brandt}, {Gallagher}, \&
  {Garmire}}]{2002ApJ...579..169C}
{Chartas}, G., {Brandt}, W.~N., {Gallagher}, S.~C., \& {Garmire}, G.~P. 2002,
  \apj, 579, 169. \eprint{arXiv:astro-ph/0207196}

\bibitem[{{Chartas} et~al.(2009){Chartas}, {Saez}, {Brandt}, {Giustini}, \&
  {Garmire}}]{2009ApJ...706..644C}
{Chartas}, G., {Saez}, C., {Brandt}, W.~N., {Giustini}, M., \& {Garmire}, G.~P.
  2009, \apj, 706, 644. \eprint{0910.0021}

\bibitem[{{Di Matteo} et~al.(2005){Di Matteo}, {Springel}, \&
  {Hernquist}}]{2005Natur.433..604D}
{Di Matteo}, T., {Springel}, V., \& {Hernquist}, L. 2005, \nat, 433, 604.
  \eprint{arXiv:astro-ph/0502199}

\bibitem[{{Ganguly} \& {Brotherton}(2008)}]{2008ApJ...672..102G}
{Ganguly}, R., \& {Brotherton}, M.~S. 2008, \apj, 672, 102. \eprint{0710.0588}

\bibitem[{{Giustini} et~al.(2011){Giustini}, {Cappi}, {Chartas}, {Dadina},
  {Eracleous}, {Ponti}, {Proga}, {Tombesi}, {Vignali}, \&
  {Palumbo}}]{2011A&A...536A..49G}
{Giustini}, M., {Cappi}, M., {Chartas}, G., {Dadina}, M., {Eracleous}, M.,
  {Ponti}, G., {Proga}, D., {Tombesi}, F., {Vignali}, C., \& {Palumbo},
  G.~G.~C. 2011, \aap, 536, A49. \eprint{1109.6026}

\bibitem[{{Knigge} et~al.(2008){Knigge}, {Scaringi}, {Goad}, \&
  {Cottis}}]{2008MNRAS.386.1426K}
{Knigge}, C., {Scaringi}, S., {Goad}, M.~R., \& {Cottis}, C.~E. 2008, \mnras,
  386, 1426. \eprint{0802.3697}

\bibitem[{{Komossa} \& {Meerschweinchen}(2000)}]{2000A&A...354..411K}
{Komossa}, S., \& {Meerschweinchen}, J. 2000, \aap, 354, 411.
  \eprint{arXiv:astro-ph/9911429}

\bibitem[{{Lanzuisi} et~al.(2011){Lanzuisi}, {Giustini}, {Cappi}, {Dadina},
  {Malaguti}, \& {Vignali}}]{2011xru..conf..240L}
{Lanzuisi}, G., {Giustini}, M., {Cappi}, M., {Dadina}, M., {Malaguti}, G., \&
  {Vignali}, C. 2011, in The X-ray Universe 2011, edited by {J.-U.~Ness \&
  M.~Ehle}, 240

\bibitem[{{McKernan} et~al.(2007){McKernan}, {Yaqoob}, \&
  {Reynolds}}]{2007MNRAS.379.1359M}
{McKernan}, B., {Yaqoob}, T., \& {Reynolds}, C.~S. 2007, \mnras, 379, 1359.
  \eprint{0705.2542}

\bibitem[{{Misawa} et~al.(2010){Misawa}, {Kawabata}, {Eracleous}, {Charlton},
  \& {Kashikawa}}]{2010ApJ...719.1890M}
{Misawa}, T., {Kawabata}, K.~S., {Eracleous}, M., {Charlton}, J.~C., \&
  {Kashikawa}, N. 2010, \apj, 719, 1890. \eprint{1007.0448}

\bibitem[{{Piconcelli} et~al.(2005){Piconcelli}, {Jimenez-Bail{\'o}n},
  {Guainazzi}, {Schartel}, {Rodr{\'{\i}}guez-Pascual}, \&
  {Santos-Lle{\'o}}}]{2005A&A...432...15P}
{Piconcelli}, E., {Jimenez-Bail{\'o}n}, E., {Guainazzi}, M., {Schartel}, N.,
  {Rodr{\'{\i}}guez-Pascual}, P.~M., \& {Santos-Lle{\'o}}, M. 2005, \aap, 432,
  15. \eprint{arXiv:astro-ph/0411051}

\bibitem[{{Proga} \& {Kallman}(2004)}]{2004ApJ...616..688P}
{Proga}, D., \& {Kallman}, T.~R. 2004, \apj, 616, 688.
  \eprint{arXiv:astro-ph/0408293}

\bibitem[{{Silk} \& {Rees}(1998)}]{1998A&A...331L...1S}
{Silk}, J., \& {Rees}, M.~J. 1998, \aap, 331, L1.
  \eprint{arXiv:astro-ph/9801013}

\bibitem[{{Sim} et~al.(2010){Sim}, {Proga}, {Miller}, {Long}, \&
  {Turner}}]{2010MNRAS.408.1396S}
{Sim}, S.~A., {Proga}, D., {Miller}, L., {Long}, K.~S., \& {Turner}, T.~J.
  2010, \mnras, 408, 1396. \eprint{1006.3449}

\bibitem[{{Tombesi} et~al.(2010){Tombesi}, {Cappi}, {Reeves}, {Palumbo},
  {Yaqoob}, {Braito}, \& {Dadina}}]{2010A&A...521A..57T}
{Tombesi}, F., {Cappi}, M., {Reeves}, J.~N., {Palumbo}, G.~G.~C., {Yaqoob}, T.,
  {Braito}, V., \& {Dadina}, M. 2010, \aap, 521, A57. \eprint{1006.2858}

\bibitem[{{Wang} et~al.(1999){Wang}, {Brinkmann}, {Wamsteker}, {Yuan}, \&
  {Wang}}]{1999MNRAS.307..821W}
{Wang}, T.~G., {Brinkmann}, W., {Wamsteker}, W., {Yuan}, W., \& {Wang}, J.~X.
  1999, \mnras, 307, 821. \eprint{arXiv:astro-ph/9903428}

\end{thebibliography}


\end{document}